\documentclass[slac_one]{revtex4}
\usepackage{graphicx}
\usepackage{fancyhdr}
\usepackage{rotating}
\usepackage{psfrag}
\newcommand{\sub}[1]{_{\mbox{\scriptsize #1}}}

\begin{document}

\title{Black Holes at Accelerators}
\author{Bryan Webber}
\affiliation{Cavendish Laboratory, University of Cambridge,
Cambridge CB3 0HE, UK}

\begin{abstract}
In theories with large extra dimensions and TeV-scale gravity,
black holes are copiously produced in particle collisions at
energies well above the Planck scale.  I briefly review some
recent work on the phenomenology of this process, with
emphasis on theoretical uncertainties and possible strategies
for measuring the number of extra dimensions.
\end{abstract}

\maketitle

\thispagestyle{fancy}

\section{INTRODUCTION}
One of the most surprising and exciting possibilities for new physics at
future colliders is the production of microscopic black
holes~\cite{Giddings:2001bu,Dimopoulos:2001hw}, which can occur in
theories with large extra dimensions\footnote{This includes warped
extra-dimension theories~\cite{Randall:1999ee} with warp scale large
compared to the black hole.}~\cite{Antoniadis:1990ew}.
String theory suggests a scenario in
which the Standard Model particles are confined to a submanifold with
the familiar three spatial dimensions (the `brane'), while gravity is free
to propagate in the full $(3+n)$-dimensional space (the `bulk').
In that case the fundamental $(4+n)$-dimensional Planck scale could
be much lower than its effective 4-dimensional value, possibly
of the order of TeV.  Then gravity would look weak on distance
scales large compared to the size of the extra dimensions,
but could become as strong as the other forces at short distances.
Parton-parton collisions with centre-of-mass energies
well above this scale could be treated as $(4+n)$-dimensional classical
gravitational interactions.  Numerical studies suggest that in such
circumstances the probability of coalescence to form a microscopic
black hole would be large~\cite{Yoshino:2002tx,Yoshino:2005hi}.
The black hole would be a powerful source
of Hawking radiation, decaying rapidly into all kinds of Standard Model
particles. From a study of the decay products, it might be possible to
deduce the number and size of the extra dimensions.

In this paper I shall review some recent work on the above scenario.
We shall see that, even if one accepts the basic assumption of TeV-scale
gravity in extra dimensions, there are many uncertainties and gaps in our
understanding, which make reliable predictions impossible at present.
Nevertheless one can build models and simulations that allow one to explore
different options for filling in the gaps, and enable experimentalists
to start thinking about how they might analyse black hole events.

I apologise for not mentioning all relevant topics and papers in this
short review.  For more complete discussion and references, please
see \cite{Harris:2004xt} and Chris Harris's thesis~\cite{Harris:2004mf}.

\section{BLACK HOLE PRODUCTION}
On purely dimensional grounds we expect the parton-level cross section
for black hole formation to be of the form
\begin{equation}
\label{fndef}
\hat\sigma(\hat s=M\sub{BH}^2)=F_n\pi r\sub{S}^2\,.
\end{equation}
where $r\sub{S}$ is the Schwarzschild radius in $(4+n)$ dimensions,
\begin{equation}
\label{rsdef}
r\sub{S}=\frac 1{\sqrt\pi M\sub{PL}}
\left[\frac{8\Gamma\left(\frac{n+3}2\right)M\sub{BH}}
{(n+2)M\sub{PL}}\right]^{\frac 1{n+1}}
\end{equation}
and $F_n$ is a ``formation factor'' of order unity. Notice that
we assume here that the black hole mass $M\sub{BH}$ is equal to
the full parton centre-of-mass energy $\sqrt{\hat s}$; this point will
be discussed further later.
We use the Dimopoulos--Landsberg~\cite{Dimopoulos:2001hw}
definition of the Planck mass,
\begin{equation}
\label{mpldef}
M\sub{PL}=
\left[G_{(4+n)}\right]^{-\frac{1}{n+2}}
\end{equation}
where $G_{(4+n)}$ is the $(4+n)$-dimensional Newton constant.
For illustrative purposes,
we shall usually take $M\sub{PL}=1$ TeV in this study.

\subsection{Black Hole Formation Factor}
\begin{figure}
 \includegraphics[width=0.65\textwidth]{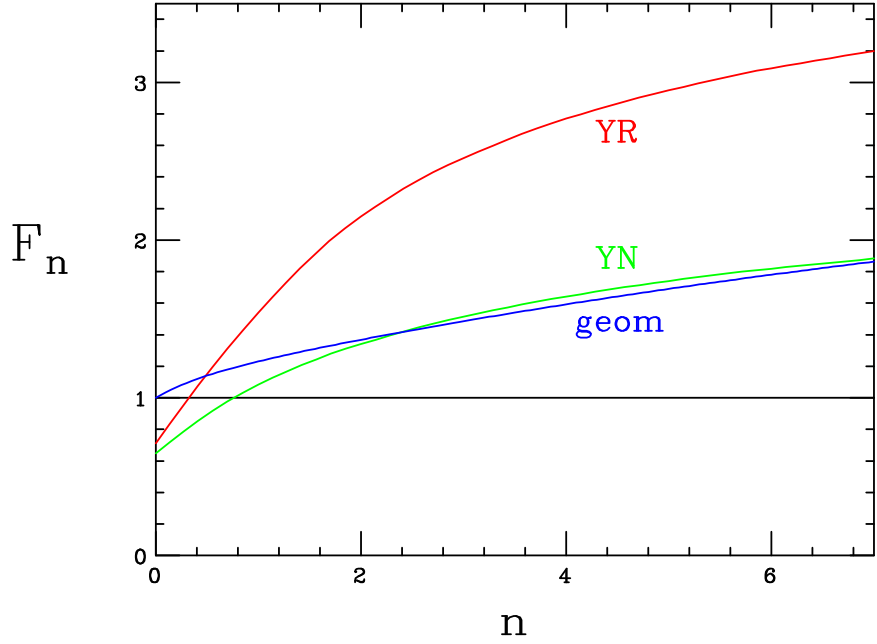}%
 \caption{Yoshino-Rychkov (YR), Yoshino-Nambu (YN) and geometrical (geom)
estimates of black hole formation factor.\label{fig:bhff}}
 \end{figure}

The formation factor $F_n$ in Equation~\ref{fndef} has been estimated in a
variety of ways.  A simple geometric argument~\cite{Ida:2002ez} goes as
follows.  Consider incoming partons that would pass
each other at a separation (impact parameter) $b=2r_h$ where $r_h$ is
the horizon radius for a Kerr black hole with the corresponding angular
momentum $J=b\sqrt s/2$ and mass $M\sub{BH}=\sqrt s$.  This is assumed to be
the maximum impact parameter at which black hole formation could occur. Therefore
\begin{equation}
\label{hatsig}
\hat\sigma = F_n\pi r\sub{S}^2 \sim \pi (2r_h)^2\;.
\end{equation}
But for a Kerr black hole in $(4+n)$ dimensions
\begin{equation}
\label{rhdef}
r_h = r\sub{S}\left[1+a_*^2\right]^{-\frac 1{n+1}}
\end{equation}
where
\begin{equation}
\label{astdef}
a_*= \frac{(n+2)J}{2r_h M\sub{BH}}
\end{equation}
and so we obtain
\begin{equation}
\label{Fgeom}
F_n\sim 4\left[1+\left(\frac{n+2}2\right)^2\right]^{-\frac 2{n+1}}
\qquad\mbox{(``geometric'').}
\end{equation}
This formula, shown by the blue curve in Figure~\ref{fig:bhff}, follows
quite closely the numerical estimate of Yoshino and Nambu~\cite{Yoshino:2002tx}
(green). On the other hand, the latter is only a lower bound on the
cross section, obtained by finding a closed trapped surface on a
particular slice of $(4+n)$-dimensional space-time~\cite{Eardley:2002re};
such a surface must be shielded by an event horizon. In this way one
obtains a lower bound on the impact parameter for horizon formation,
and hence on the cross section for black hole formation.

More recently, Yoshino and Rychkov~\cite{Yoshino:2005hi} have found a
more optimal space-time slice, which leads to a larger lower bound on the
formation factor, shown in red in Figure~\ref{fig:bhff}.  

\subsection{Black Hole Cross Section}
\begin{figure}
 \includegraphics[width=0.65\textwidth]{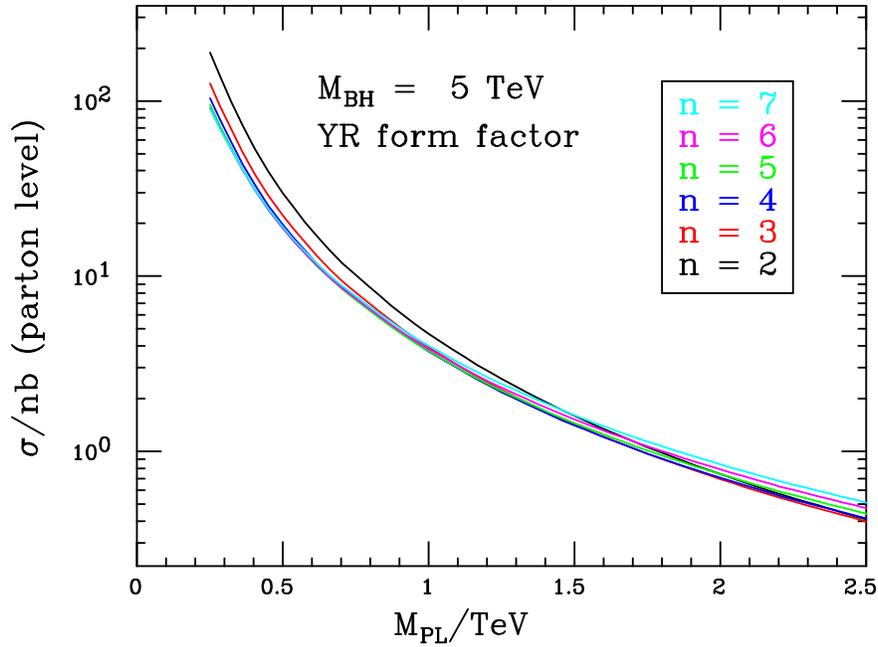}%
 \caption{Parton-level black hole cross section.\label{fig:bhcs}}
 \end{figure}

Adopting the Yoshino-Rychkov lower bound as an estimate of the black hole formation
factor, and assuming again that $\sqrt{\hat s}=M\sub{BH}$,
we obtain the parton-level cross section, shown as a function of the Planck
scale for a 5 TeV black hole in Figure~\ref{fig:bhcs}. For $n>2$ extra dimensions,
the $n$-dependence of the formation factor tends to cancel that of the Schwarzschild
radius, so that the cross section is not strongly dependent on the number of extra
dimensions.\footnote{The same is true for the Yoshino-Nambu estimate, only the numerical
values are smaller.} Values of $n$ less than 3 are in any case strongly disfavoured
on astrophysical grounds~\cite{Hannestad:2001xi}.

 \begin{figure}
 \includegraphics[width=0.65\textwidth]{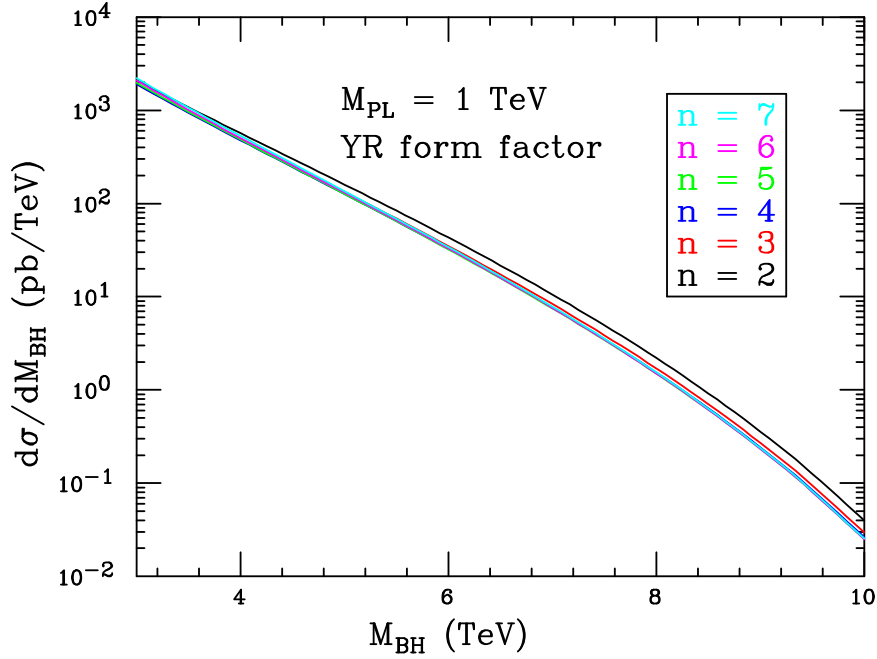}%
 \caption{Hadron-level black hole cross section at the LHC.\label{fig:bhLHC}}
 \end{figure}

To estimate the cross section for black hole production at a hadron collider,
we must convolve the parton-level cross section with the parton distributions in
the incident hadrons. The resulting cross section for a $pp$ collider at c.m.\ energy
14 TeV (i.e.\ the LHC) is shown in Figure~\ref{fig:bhLHC}.  At the LHC design luminosity
of $10^{34}$ cm$^{-2}$s$^{-1}$, this corresponds to more than one black hole per second
with mass above 5 TeV.

\subsection{Measuring the Planck Scale}
If we accept the Yoshino-Rychkov result as a reliable estimate of the black hole formation
factor, we see from Figures~\ref{fig:bhcs} and \ref{fig:bhLHC} that a measurement of the
cross section for a given range of black hole masses would fix the Planck mass in a way
that is substantially independent of the number of extra dimensions, at least in the
astrophysically favoured region $n>2$.

Of course, one expects some dramatic changes in the cross section and final
state at partonic c.m.\ energies around the Planck scale, due to the onset of
strong gravitational scattering.  However, to predict those changes one would need
a quantum theory of gravity, so deducing the Planck mass from them would not
be straightforward. Measurements well above the Planck scale, on the other hand,
can reasonably be interpreted in a classical approximation, as we are doing here.

The problem in any case is to make a reliable measurement of the black hole mass, or
more correctly the partonic c.m.\ energy for black hole formation.  Since we do not
observe the colliding partons, this can only be inferred from properties of the final
state, which will be dominated by the decay of the black hole.

\section{BLACK HOLE DECAY}
Although the formation of a horizon in parton collisions well above the Planck scale
seems reliably established, the nature and fate of the object thus created is much
less clear.  The usual working hypothesis has been that the evolution of the system
has four phases:
\begin{itemize}
\item{\em Balding phase}: all `hair' (characteristics other than mass, charge and
angular momentum) and multipole moments are lost through gravitational and Hawking
radiation, and the object becomes the multidimensional generalization of a
Kerr-Newman black hole.  In fact any residual charge after this phase is
probably negligible~\cite{Giddings:2001bu}, so the Kerr solution is assumed.
\item{\em Spin-down phase}: the Kerr black hole loses angular momentum by
Hawking radiation and becomes a Schwarzschild black hole.
\item{\em Schwarzschild phase}: the black hole loses mass through
Hawking radiation and its temperature rises until the mass and/or
temperature reach the Planck scale.
\item{\em Planck phase}: the object (`string ball'?) is in the realm of quantum gravity
and its fate cannot be predicted.  It could decay into a few quanta with Planck-scale
energies~\cite{Giddings:2001bu}, evaporate at the Hagedorn
temperature~\cite{Dimopoulos:2001qe}, or even form a new kind
of stable relic object~\cite{Koch:2005ks}.
\end{itemize}
At present, each of these decay phases is subject to great uncertainty.
The amount of gravitational radiation emitted in the balding phase
is of major concern because this constitutes missing energy that
would spoil the connection between collision energy and black hole
mass\footnote{Lower bounds on the black hole mass can be deduced from
the trapped surface area~\cite{Eardley:2002re,Yoshino:2005hi}.}~\cite{Yoshino:2005hi,
Anchordoqui:2003ug,Cardoso:2005jq}.
In higher dimensions there are solutions to the Einstein equation that
are not simply generalizations of the four-dimensional solutions, such
as black rings~\cite{Horowitz:2005rs}.
Probably there are more complicated objects still to be
discovered.  It is not clear whether such configurations would be
able to spin down to the Schwarzschild solution, or what their
Hawking radiation would look like.  Even assuming the generalized
Kerr solution, the amount and distribution of Hawking radiation during
spin-down is still under
investigation~\cite{Harris:2005jx,Duffy:2005ns,Casals:2005sa,Ida:2006tf}.

In the Schwarzschild phase too there are many points that require
further clarification. What fraction of the Hawking radiation is emitted
as detectable Standard Model particles on the brane, and how much
escapes into the bulk?  Is the decay process too rapid for the relationship
between black hole mass and temperature to remain valid throughout?
Do secondary decays significantly distort the Hawking spectrum?
And in the final Planck phase, what are the consequences of
alternative models for the fate of the remnant object?

Many of these questions await theoretical answers, but some can be illuminated
by numerical simulations.  For the latter, we shall start with the hypotheses
that practically all the energy of the parton collision goes into the black
hole, that the decay is dominated by the Schwarzschild phase, and that the Hawking
radiation consists entirely of Standard Model particles on the brane.  With
these assumptions, a detailed picture of the final state can be presented
and the effects of some of the uncertainties can be investigated.

\subsection{Hawking Spectrum}
With the above assumptions, the spectrum of particles emitted during black hole
decay takes the form
\begin{equation}
\label{spectrum}
\frac{dN}{dE}\propto
\frac{\gamma E^2}{(e^{E/T\sub{H}}\mp 1)}T\sub{H}^{n+6}
\end{equation}
where as usual the $\mp$ applies to bosons and fermions,
$T\sub{H}$ is the Hawking temperature
\begin{equation}
\label{temp}
T\sub{H} = \frac{n+1}{4\pi r\sub{S}}\propto M\sub{BH}^{-\frac 1{n+1}}
\end{equation}
and $\gamma$ is a $(4+n)$-dimensional {\em grey-body factor}
\cite{Harris:2004mf,Ida:2002ez,Kanti:2002nr,Harris:2003eg}.
The latter takes account of the fact that the wave function of a particle
created in the intense gravitational field near the horizon has to
propagate through the curved space around the black hole in order
for the particle to be observed.  By detailed balance, the grey-body
factor is equal to the absorption coefficient of the
black hole for waves incident from infinity.
For particles emitted at high energies the wavenumber is large compared
with the curvature, so grey-body effects are small and the spectrum
is close to black-body.  But at low energy the emission is strongly
modified in a way that depends on the spin and the number of extra
dimensions.

\begin{figure}
\psfrag{x}[][][2.5]{$E\,r\sub{S}$}
\psfrag{y}[][][2.5]{$\hat{\sigma}^{(0)}\sub{abs}/\pi r\sub{S}^2$}
\scalebox{0.4}{\rotatebox{-90}{\includegraphics[width=\textwidth]{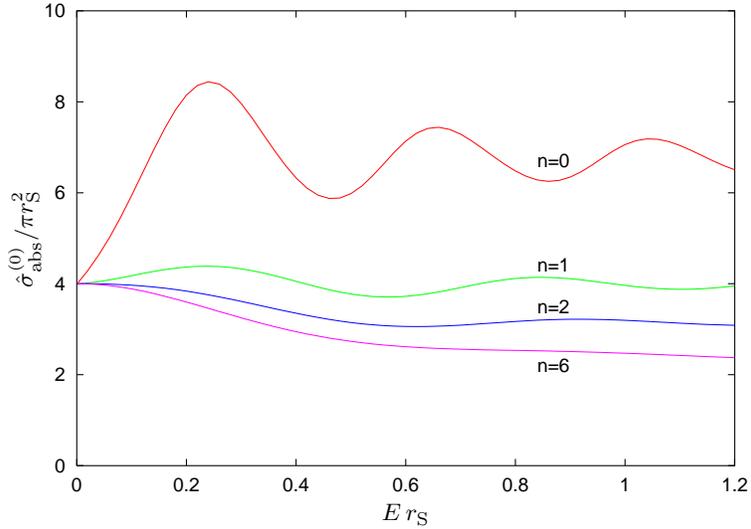}}}%
\caption{Grey-body factors for scalar emission on the brane from a $(4+n)$D
black hole.\label{grey0}}
\end{figure}

\begin{figure}
\psfrag{x}[][][2.5]{$E\,r\sub{S}$}
\psfrag{y}[][][2.5]{$\hat{\sigma}^{(1/2)}\sub{abs}/\pi r\sub{S}^2$}
\scalebox{0.4}{\rotatebox{-90}{\includegraphics[width=\textwidth]{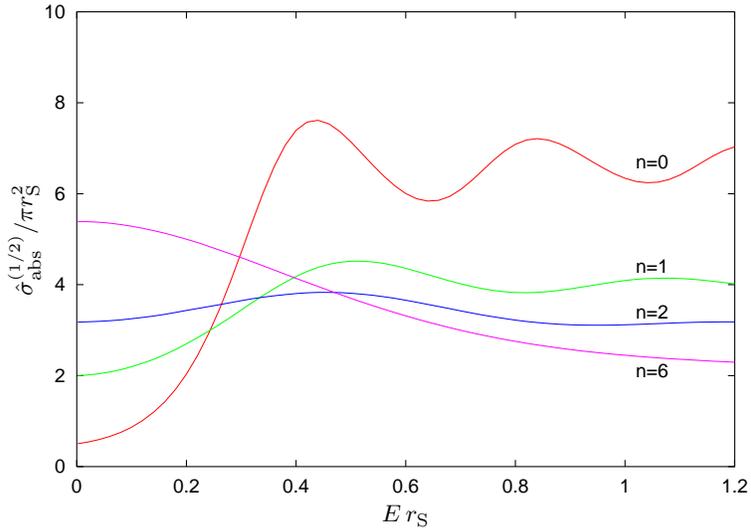}}}%
\caption{Grey-body factors for fermion emission on the brane from a $(4+n)$D
black hole.\label{grey05}}
\end{figure}

\begin{figure}
\psfrag{x}[][][2.5]{$E\,r\sub{S}$}
\psfrag{y}[][][2.5]{$\hat{\sigma}^{(1)}\sub{abs}/\pi r\sub{S}^2$}
\scalebox{0.4}{\rotatebox{-90}{\includegraphics[width=\textwidth]{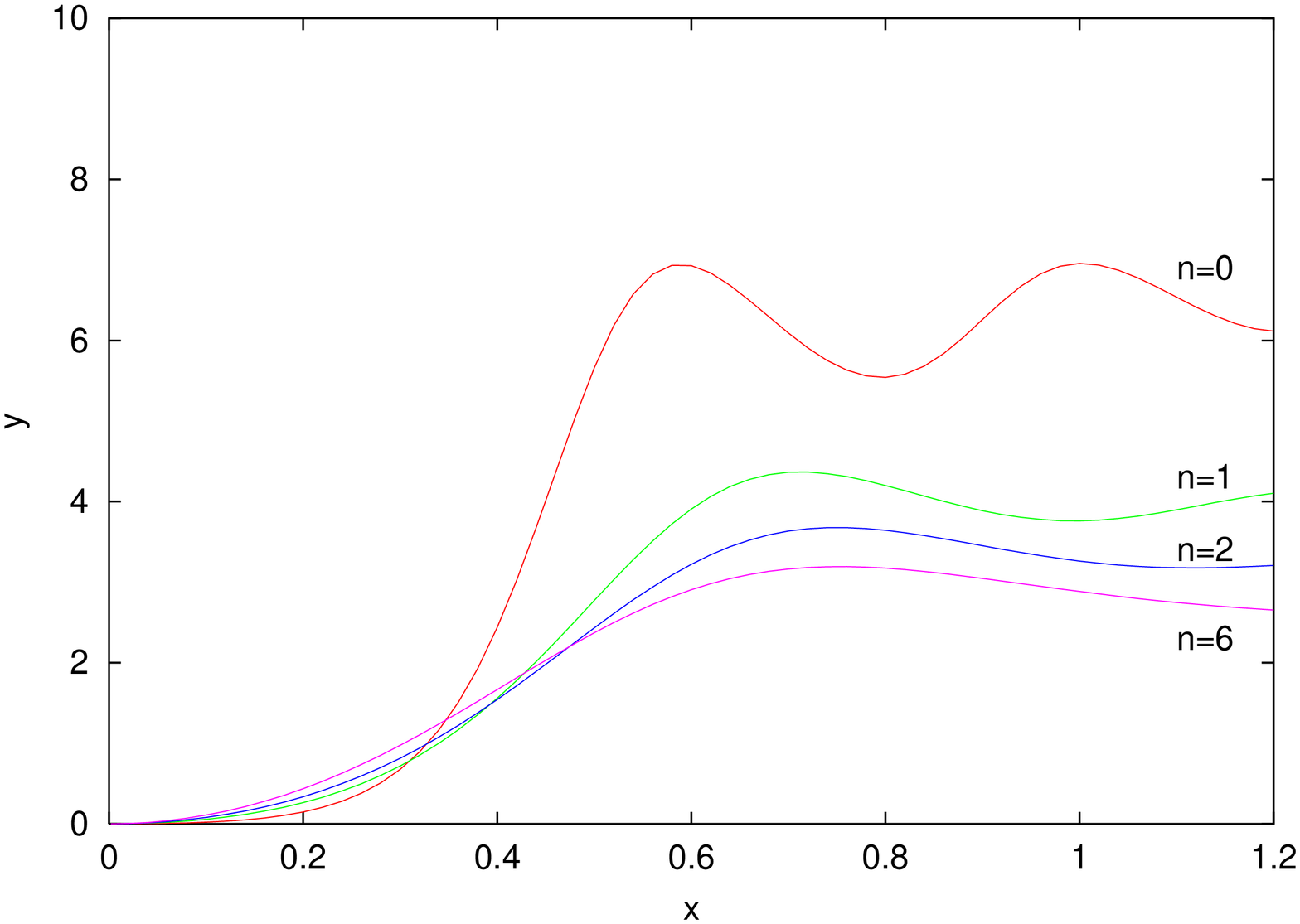}}}%
\caption{Grey-body factors for gauge boson emission on the brane from a $(4+n)$D
black hole.\label{grey1}}
\end{figure}

Figures~\ref{grey0}--\ref{grey1}, taken from \cite{Harris:2004mf}, show the
grey-body factors for scalars, fermions and gauge bosons as functions of
the particle energy in units of the inverse horizon radius. What is actually
plotted here is the absorption cross section in units of $\pi r\sub{S}^2$,
which tends to 4 at high energies. We see that the main effect in extra
dimensions is the suppression of low-energy gauge boson emission.

\subsection{Integrated Flux and Lifetime}
Integrating the spectrum in Equation 8, including the grey-body factors,
and multiplying by the number of Standard Model degrees of
freedom for each spin, one obtains~\cite{Harris:2004mf} the total
flux of particles of each type in black hole decay.

\begin{figure}
\psfrag{x}[][][2.5]{$n$}
\psfrag{y}[][][2.5]{$F\,r\sub{S}$}
\scalebox{0.5}{\rotatebox{-90}{\includegraphics[width=\textwidth]{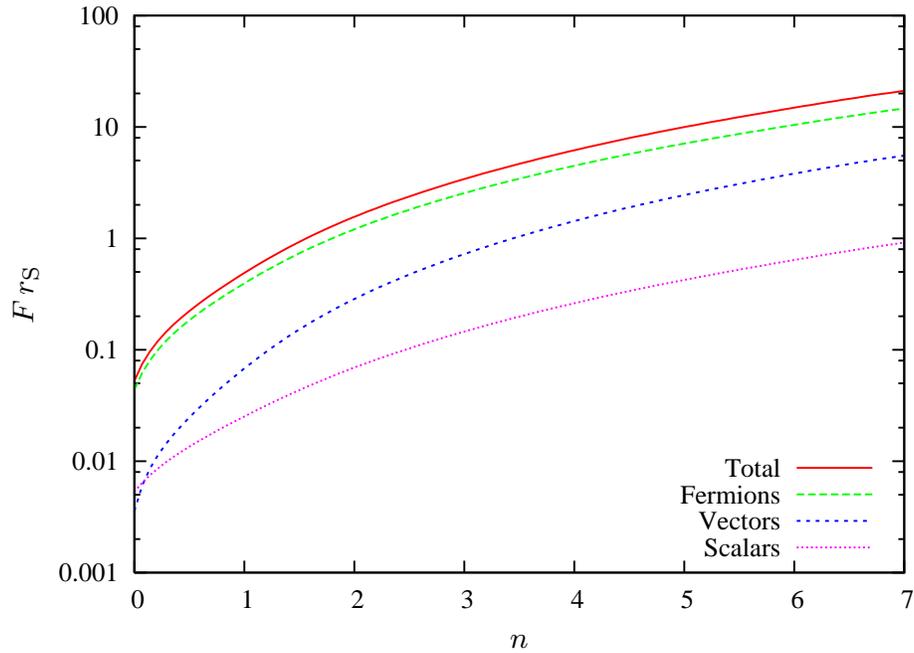}}}%
\caption{Integrated flux of Hawking emission on the brane in the decay of a $(4+n)$D
black hole.\label{flux}}
\end{figure}

Expressed in units of the inverse horizon radius, as shown in Figure~\ref{flux},
the particle fluxes are independent of the black hole
mass and Planck scale.  When the flux in these units exceeds unity,
which we see is the case for $n>2$,
the time between emissions is less than the time for a light-signal
to travel a distance equal to the horizon radius. In these circumstances
it is difficult to see how the emission can remain thermal.  However,
we shall continue to make that assumption in the absence of any better
understanding.

\begin{figure}
\psfrag{x}[][][2.5]{$n$}
\psfrag{y}[][][2.5]{$\tau\,M\sub{BH}$}
\psfrag{w}[][][1.75]{$M\sub{BH}=5M\sub{PL}$}
\psfrag{z}[][][1.75]{$M\sub{BH}=10M\sub{PL}$}
\scalebox{0.5}{\rotatebox{-90}{\includegraphics[width=\textwidth]{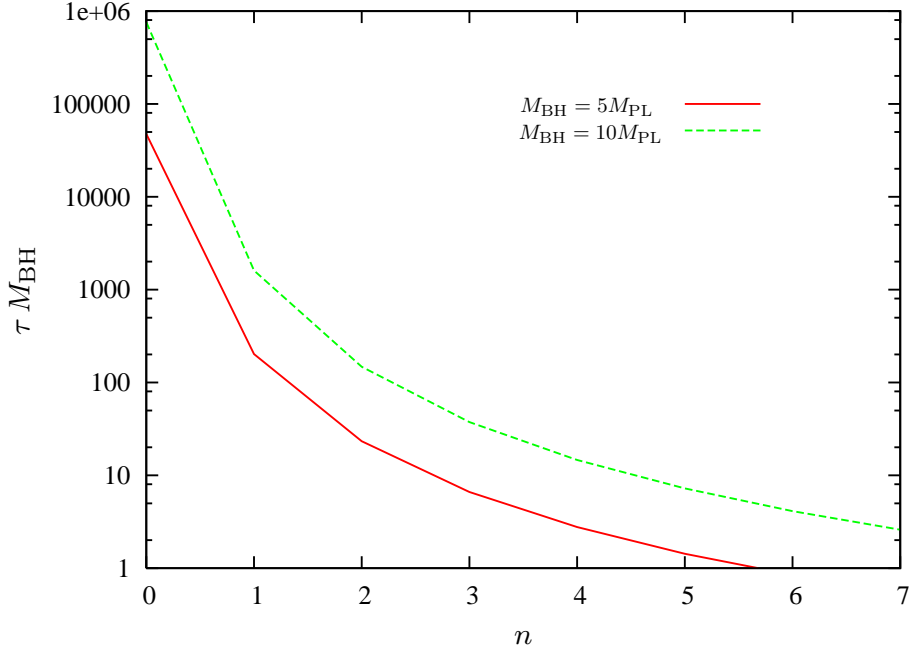}}}%
\caption{Mean lifetime of a $(4+n)$D black hole.\label{life}}
\end{figure}

Assuming that the Schwarzschild phase of decay is dominant, and that the
mass and temperature are related by Equation \ref{temp} throughout
(which we have just seen must be doubtful), the total energy flux can
be integrated~\cite{Harris:2005jx} to find the time at which the entire
mass of the black hole has been radiated away. This measure of the
lifetime, expressed in units of the inverse of the initial mass, is
shown in Figure~\ref{life}.  We see that the lifetime falls very steeply
as a function of the number of dimensions, and indeed can be comparable
with the inverse mass when $n>4$, even for masses well above the Planck scale.
When this is the case, the object formed can no longer really be said to
have an independent existence as a black hole.

\section{EVENT SIMULATION}

\subsection{CHARYBDIS Event Generator}
The simulation program {\small CHARYBDIS}~\cite{Harris:2003db} generates
black hole production and decay configurations assuming that all the partonic collision
energy goes into the mass of the hole, and that the Schwarzschild decay phase
dominates.\footnote{The current public version (1.001) neglects the form factor $F_n$
in Equation~\ref{fndef}, so that the production cross section is somewhat smaller than
it should be.}  The identities and momenta of the incoming partons and outgoing
primary decay products are passed to the {\small HERWIG}~\cite{Corcella:2000bw}
event generator via the Les Houches interface~\cite{Boos:2001cv}. {\small HERWIG}
then handles all the QCD parton showering, hadronization and secondary decays.

\begin{table}[t]
\def\arraystretch{1.1}
\begin{center}
\begin{tabular}{|c|l|c|c|}
\hline
Name & Description & Values & Default\\
\hline
\texttt{MINMSS} & Minimum mass of black holes (GeV) & $<\mbox{\texttt{MAXMSS}}$ & \texttt{5000.0}\\
\texttt{MAXMSS} & Maximum mass of black holes (GeV) & $\leq\mbox{c.m. energy}$ & c.m\ energy\\
\texttt{MPLNCK} & Planck mass (GeV) & $\leq\mbox{\texttt{MINMSS}}$ & \texttt{1000.0}\\
\texttt{TOTDIM} & Total number of dimensions ($4+n$) & \texttt{6}--\texttt{11} & \texttt{6}\\
\texttt{TIMVAR} & Allow $T\sub{H}$ to change with time & \texttt{LOGICAL}& \texttt{.TRUE.}\\
\texttt{MSSDEC} & Choice of decay products & \texttt{1}--\texttt{3} & \texttt{3}\\
\texttt{GRYBDY} & Include grey-body effects & \texttt{LOGICAL} & \texttt{.TRUE.}\\
\texttt{KINCUT} & Use a kinematic cut-off on the decay & \texttt{LOGICAL} & \texttt{.FALSE.}\\
\texttt{NBODY} & Number of particles in remnant decay & \texttt{2}--\texttt{5} & \texttt{2}\\
\hline
\end{tabular}
\caption{Main {\small CHARYBDIS} parameters.\label{parameters}}
\end{center}
\def\arraystretch{1.0}
\end{table}

The main parameters that control the operation of {\small CHARYBDIS} are summarized
in Table~\ref{parameters}.  The first four are self-evident. The remaining five
provide the means to study the effects of some of the uncertainties discussed
earlier.  \texttt{TIMVAR} causes the
temperature of the black hole to be updated according to Equation~\ref{temp} after
each emission; otherwise it is frozen at the initial value. \texttt{MSSDEC} controls
whether heavy particles are included in the Hawking radiation:  \texttt{MSSDEC}=1
allows only light particle emission (up to and including $b$ quarks);
\texttt{MSSDEC}=2 includes top quark, W and Z emission; \texttt{MSSDEC}=3
includes also Higgs boson emission.

The parameters \texttt{KINCUT} and \texttt{NBODY} determine how the
evolution of the black hole is terminated.
If \texttt{KINCUT=.TRUE.}, termination occurs when the chosen energy for an emitted
particle is ruled out by the kinematics of a two-body decay.  At this point an
isotropic \texttt{NBODY} decay is performed on the black hole remnant.
The \texttt{NBODY} particles are chosen according to the same probabilities
used for the first part of the decay. The selection is then accepted if charge
and baryon number are conserved, otherwise a new set of particles is picked for
the decay.
In the alternative termination (\texttt{KINCUT=.FALSE.}) particles are emitted
according to their Hawking energy spectra until $M\sub{BH}$ falls below
\texttt{MPLNCK}; then an \texttt{NBODY} decay as described above is performed.
Any chosen energies which are kinematically forbidden are simply discarded.

\begin{figure}
\begin{center}
\includegraphics[width=0.65\textwidth]{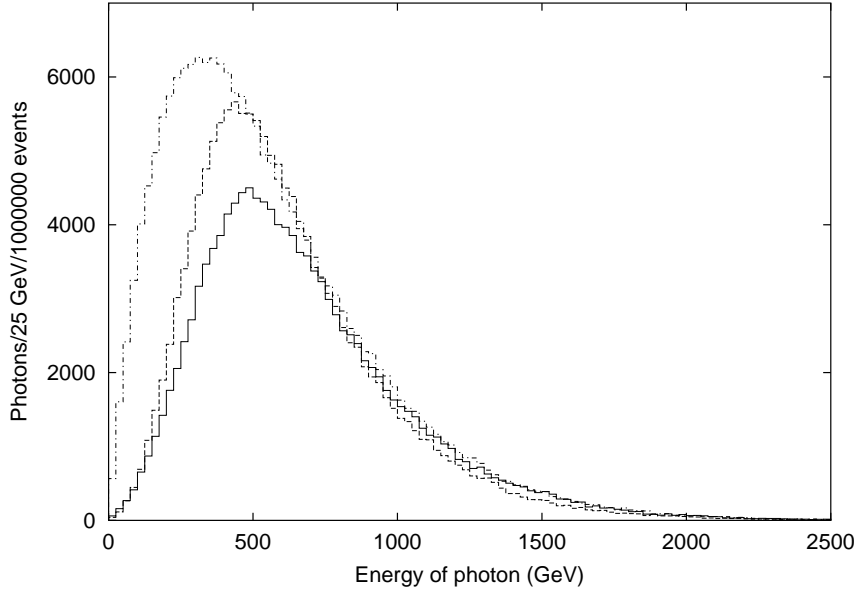}
\caption{Primary photon spectrum from black holes with initial masses 5.0--5.5 TeV in
$n=2$ extra dimensions.
 Dashed: neglecting time-variation of temperature.  Dot-dashed: neglecting grey-body factors.
Solid: including both.\label{photon}}
\end{center}
\end{figure}
 
Figure~\ref{photon} illustrates the effects of grey-body factors and temperature
variation on the energy distribution of primary photons.  Both tend to harden the
spectrum, shifting the peak towards higher energies.  They also reduce the total
number of photons emitted, the grey-body factor by suppressing soft photon emission
and the temperature variation by reducing the lifetime of the black hole.

\begin{table}
\begin{center}
\begin{tabular}{|l|c|c|c|c|}
\hline
& \multicolumn{4}{c|}{Particle emissivity (\%)} \\
\cline{2-5}
& \multicolumn{2}{c|}{\texttt{GRYBDY=.TRUE.}} & \multicolumn{2}{c|}{\texttt{GRYBDY=.FALSE.}}\\
\cline{2-5}
Particle type & Generator & Theory & Generator & Theory\\
\hline
Quarks & 63.9 & 61.8 & 58.2 & 56.5\\
Gluons & 11.7 & 12.2 & 16.9 & 16.8\\
Charged leptons  & 9.4 & 10.3 & 8.4 & 9.4\\
Neutrinos & 5.1 & 5.2 & 4.6 & 4.7\\
Photon & 1.5 & 1.5 & 2.1 & 2.1\\
Z$^0$ & 2.6 & 2.6 & 3.1 & 3.1\\
W$^+$ and W$^-$ & 4.7 & 5.3 & 5.7 & 6.3\\
Higgs boson & 1.1 & 1.1 & 1.0 & 1.1\\
\hline
\end{tabular}
\caption{Relative numbers of primary emissions from black holes with initial
masses 5.0--5.5 TeV in $n=2$ extra dimensions..\label{probtab}} 
\end{center}
\end{table}

Table~\ref{probtab} shows the relative numbers of primary particles of
different types emitted when the parameters \texttt{TIMVAR}, \texttt{GRYBDY},
\texttt{NBODY} and \texttt{NBODY} have their default values, compared to the
values obtained by integrating the theoretical spectra.  The minor discrepancies
are due mainly to kinematic constraints, particle masses and charge conservation.

\subsection{Event Characteristics}
\begin{figure}
\begin{center}
\includegraphics[width=0.65\textwidth]{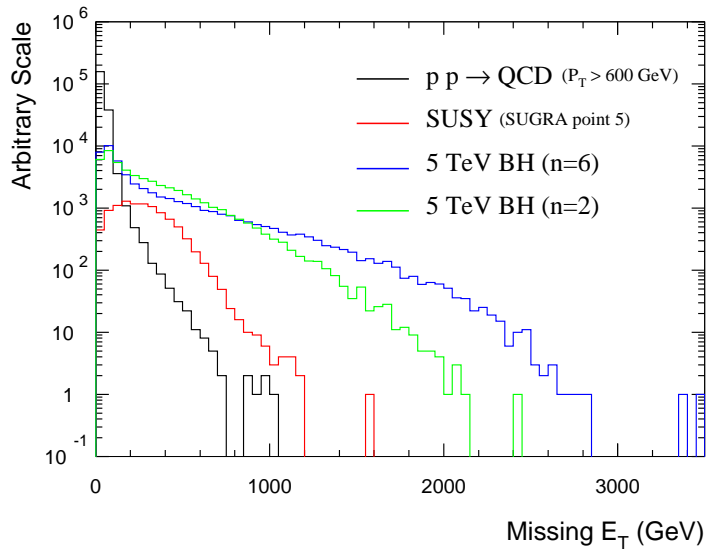}
\caption{Missing transverse energy for various processes at the LHC.\label{missEt}}
\end{center}
\end{figure}

Turning from single-particle spectra to overall event characteristics, one
interesting feature is that black hole decay is associated with large missing
transverse energy, due to copious primary and secondary neutrino emission.
Figure~\ref{missEt}, from \cite{Harris:2004xt}, shows a comparison with
expected QCD and SUSY missing $E_T$ distributions generated using
{\small HERWIG}~\cite{Corcella:2000bw,Moretti:2002eu}.  We see that
the black hole missing $E_T$ is typically larger even than that
in supersymmetric processes, where missing energy is mostly carried off by
a pair of neutralinos.  The effect is partly due to the larger
mass of the black hole, relative to the assumed SUSY scale.  However,
a large cross section with large missing energy would clearly be a good
initial indicator of black hole production.

\subsection{Measuring Black Hole Masses}
\begin{figure}
\begin{center}
\includegraphics[width=0.49\textwidth]{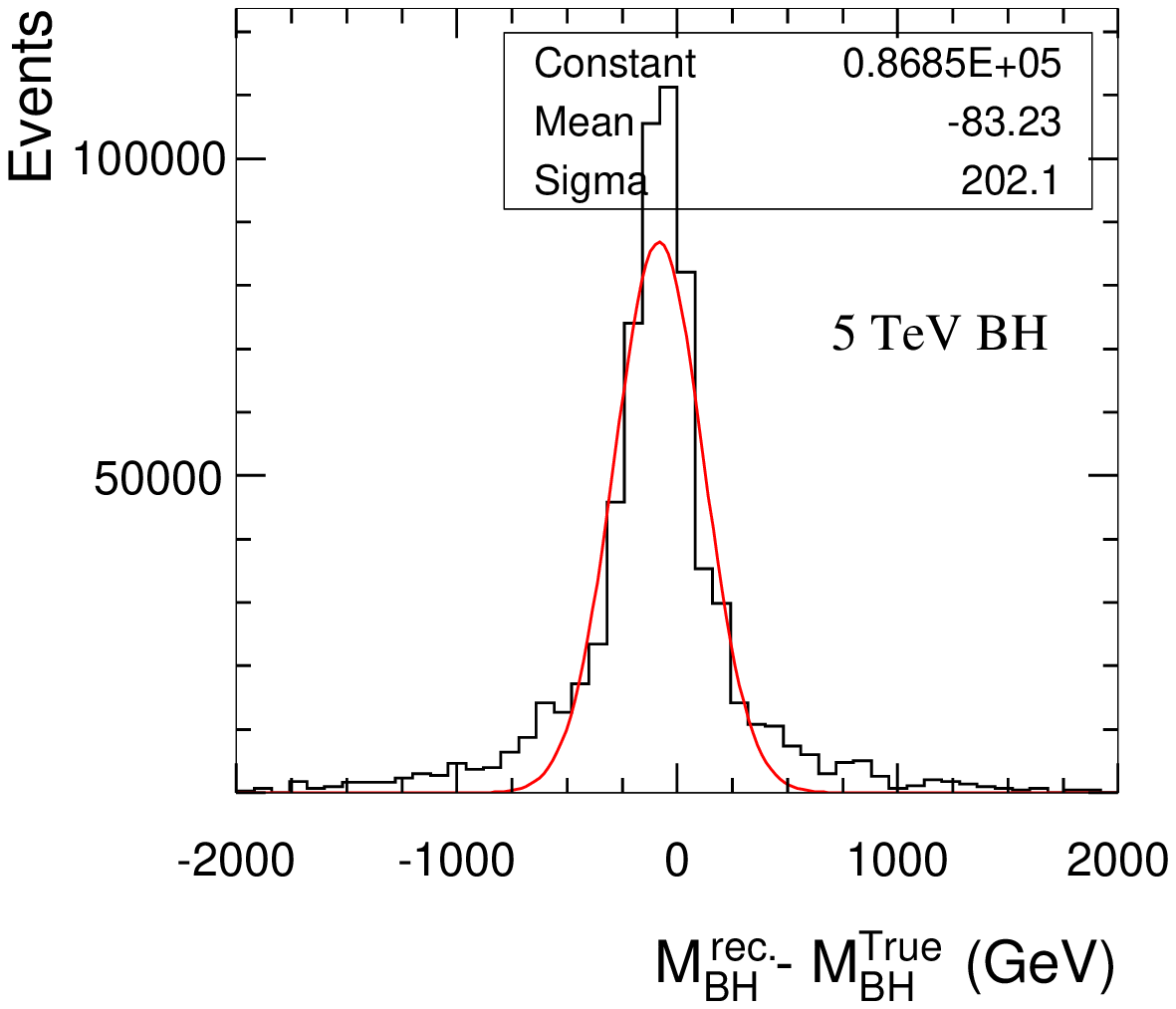}
\includegraphics[width=0.49\textwidth]{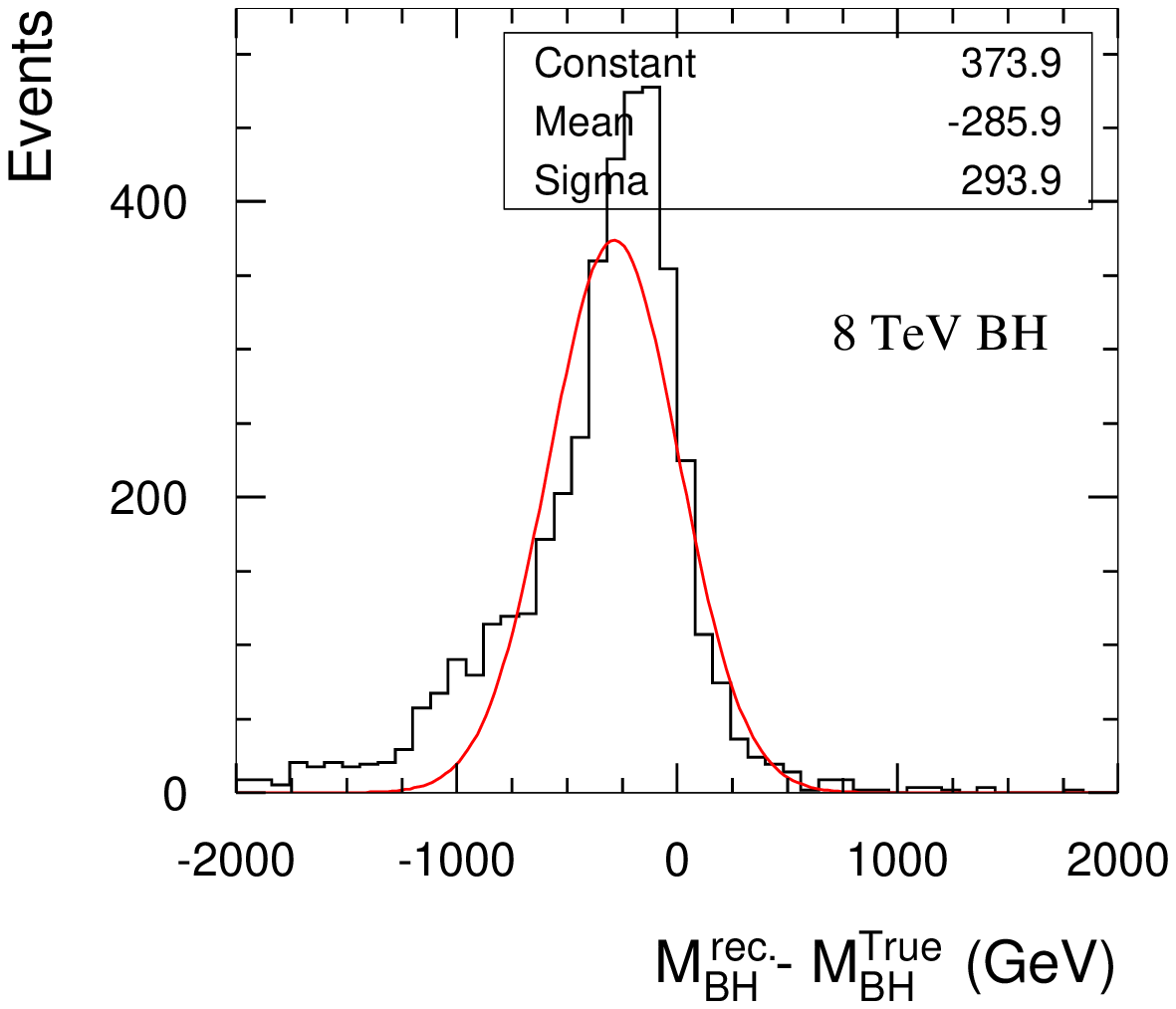}
\caption{Black hole mass resolution after cuts.\label{bhmass}}
\end{center}
\end{figure}

The large missing energy in black hole decay poses a problem for the reconstruction
of the mass of the black hole from its decay products.  In \cite{Harris:2004xt}
we found that a cut on missing $E_T<100$ GeV was necessary for a useful mass
resolution of around 4\%, i.e. $\pm 200$ GeV at 5 TeV, as illustrated in
Figure~\ref{bhmass}.  At this low value of missing $E_T$, QCD background
has to be controlled by requiring at least 4 high-$E_T$ jets. 

\section{DETERMINING THE NUMBER OF EXTRA DIMENSIONS}
\subsection{Fitting Emission Spectra}
In the event that black holes are produced at the LHC, the quantity of principal
interest will be the number of extra dimensions, $n$. Given the sensitivity of
the Hawking temperature to $n$ in Equation~\ref{temp}, it would seem
that fitting the emission spectra for various black hole masses would allow
one to extract this quantity~\cite{Dimopoulos:2001hw}.  However, we found
in~\cite{Harris:2004xt} that the theoretical uncertainties, together
with the distortion of the spectrum by secondary decays, would
make it difficult to have confidence in such a measurement. 

\begin{figure}
\begin{center}
\includegraphics[width=0.85\textwidth]{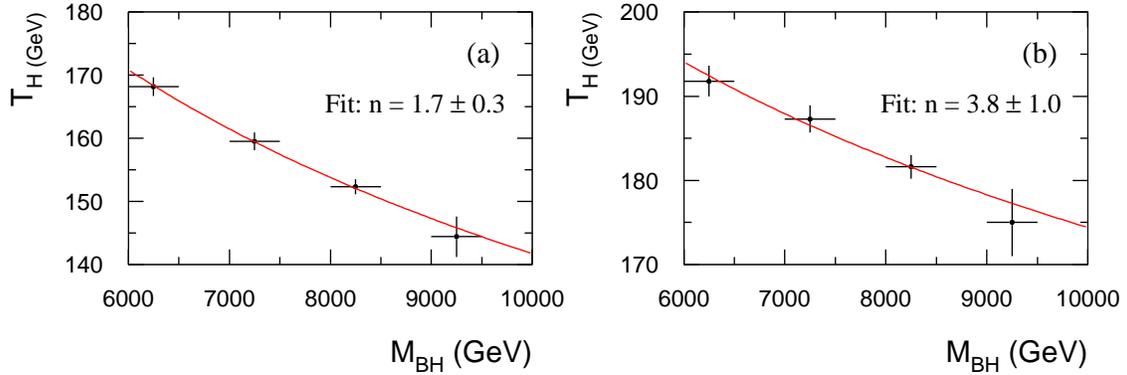}
\caption{Fits to $T\sub{H}$ vs $M\sub{BH}$ from primary electron spectrum for $n=2$.
(a) \texttt{TIMVAR} off; (b) \texttt{TIMVAR} on.\label{timvar}}
\end{center}
\end{figure}

For example, Figure~\ref{timvar} shows the effect of possible time
variation of the Hawking temperature on a fit to the primary electron spectrum
(assuming this could be unfolded cleanly from the data).  For events
generated with the temperature frozen (left), the fit gives a result
consistent with the input value $n=2$, whereas re-thermalization
between every emission (right) systematically shifts the fit
to higher values, due to the higher average temperature.  Since the
true situation would presumably lie between these extremes, the true
value of $n$ could not be extracted without a deeper understanding of
the decay process.
 
\begin{figure}
\begin{center}
\begin{minipage}{0.49\textwidth}
\includegraphics[width=\textwidth]{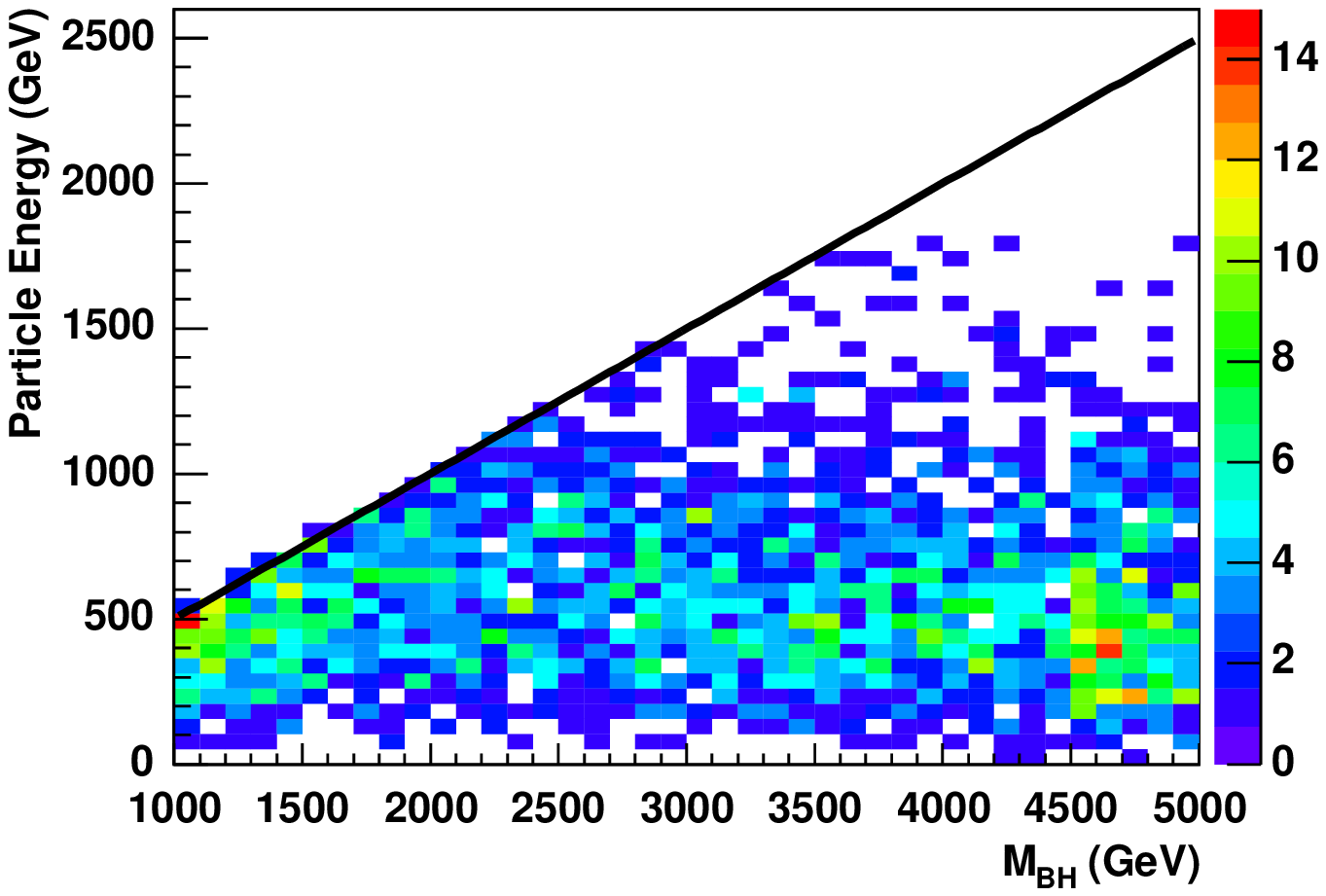}\\
(a) $n=2$.
\end{minipage}
\begin{minipage}{0.49\textwidth}
\includegraphics[width=\textwidth]{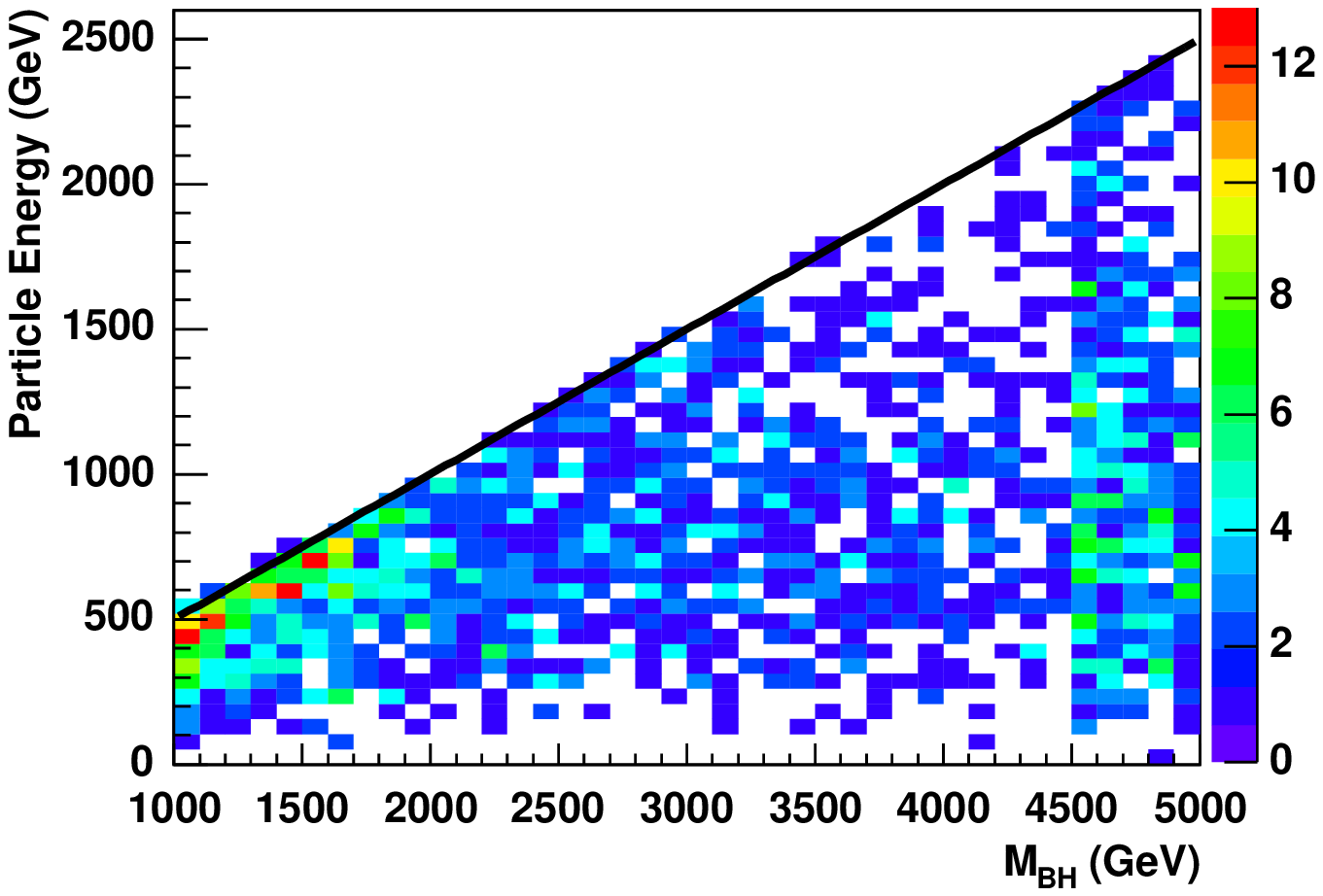}\\
(b) $n=4$.
\end{minipage}
\caption{Energy distribution of primary decay products vs $M\sub{BH}$, showing
the effect of the kinematic limit $E<M\sub{BH}/2$.
The colour code indicates the number of particles in each bin
per 1000 events.\label{cutoff}}
\end{center}
\end{figure}

Figure~\ref{cutoff} illustrates another difficulty in fitting the emission spectrum
as a function of black hole mass, this time due to kinematic effects at higher
energies.  The limitation to emission energies less than half the total mass
significantly truncates the spectrum generated according to Equation~\ref{spectrum}
at low masses and/or large values of $n$.

\subsection{A Possible Observable}
In order to avoid the low-energy region of Hawking emission, where secondary decays
distort the spectrum, and the highest energies where the kinematic cutoff takes
effect, we examined~\cite{Harris:2004xt} the region of high but not extreme
energies, $E \sim E\sub{cut} =M\sub{BH}/2 - E_d$ where $E_d$ is a few hundred GeV.
Particles in this region are also less sensitive to time variation of
the temperature, since they tend to be emitted early in the evolution
of the black hole. This tendency is further enhanced by demanding that the
highest-energy emission should have energy $E\sub{max}>E\sub{cut}$. 
We therefore looked at the fraction $F$ of events satisfing this cut
as a function of $M\sub{BH}$.

\begin{figure}
\begin{minipage}{0.49\textwidth}
\includegraphics[width=\textwidth]{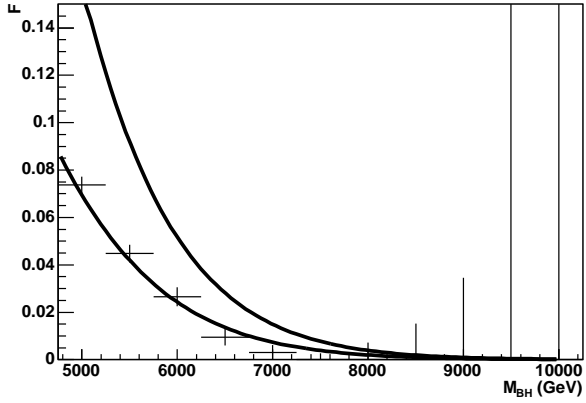}\\
(a) Default case.
\end{minipage}
\begin{minipage}{0.49\textwidth}
\includegraphics[width=\textwidth]{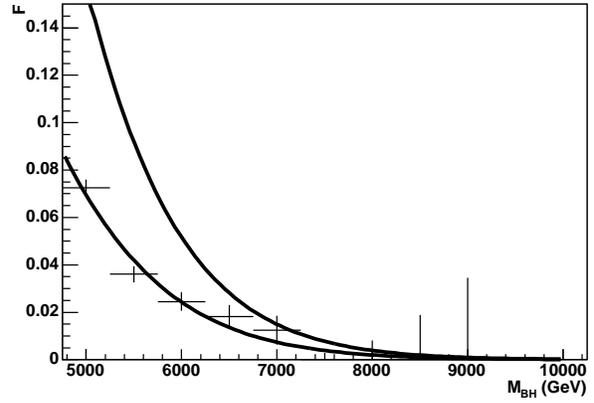}\\
(b) No time variation.
\end{minipage}
\begin{minipage}{0.49\textwidth}
\includegraphics[width=\textwidth]{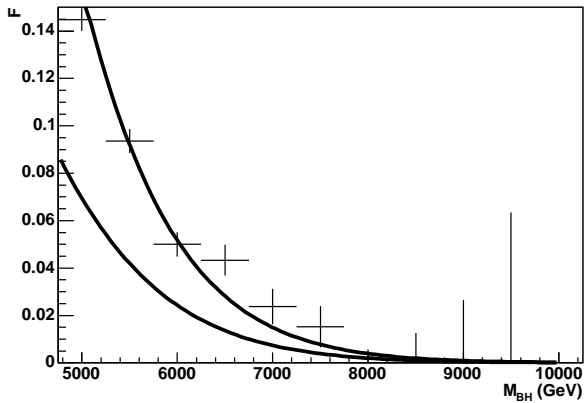}\\
(c) Kinematic cut on.
\end{minipage}
\begin{minipage}{0.49\textwidth}
\includegraphics[width=\textwidth]{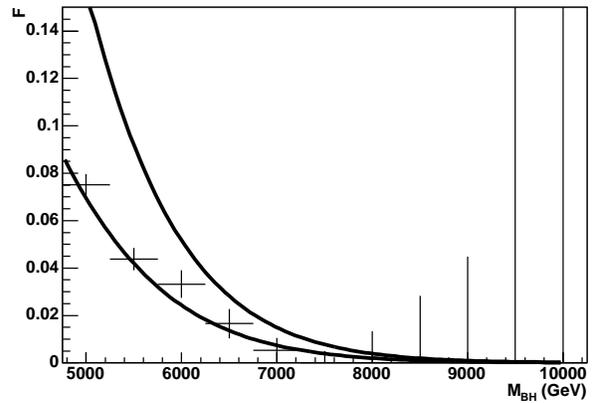}\\
(d) 4-body remnant decay.
\end{minipage}
\caption{Fraction of events with $E\sub{max}>M\sub{BH}/2 - 400$ GeV vs $M\sub{BH}$
for different options, all with $n=4$.  Upper and lower bounds
explained in the text are also shown.}
\label{fig:KLModelInvariant}
\end{figure}

As shown in Figure~\ref{fig:KLModelInvariant}, for $E_d=400$ GeV this observable is
indeed relatively insensitive to the uncertainties represented by the
{\small CHARYBDIS} parameters \texttt{TIMVAR}, \texttt{KINCUT} and
\texttt{NBODY}. The upper and lower bounds
were obtained by integrating the Planck spectrum from $E\sub{cut}$ up to infinity
and up to $M\sub{BH}/2$, respectively, and include a $\pm 200$ GeV systematic
uncertainty in the black hole mass.  The value of $F$ always lies within, or at
least is consistent with, the expected band.  Therefore a fit to this observable
should have less model dependence than analyses based, for example, on a fit to
the full emission spectrum. 

\begin{figure}
\begin{center}
\includegraphics[width=0.65\textwidth]{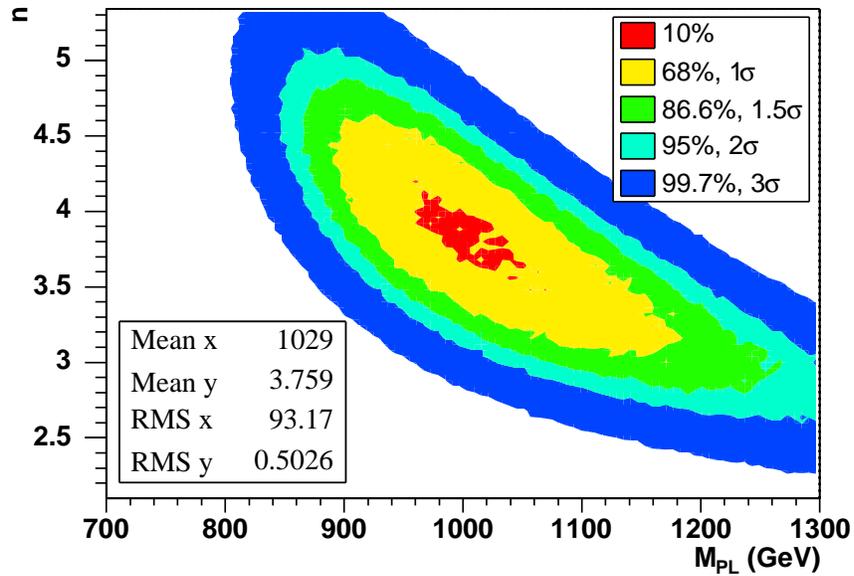}
\caption{Combined measurement of Planck mass and number of extra dimensions.\label{KL_N_MPl}}
\end{center}
\end{figure}

In~\cite{Harris:2004xt} we combined a fit to this observable and a cross-section
measurement with an assumed error of 20\%, to make a joint determination of the
Planck mass and the number of extra dimensions.  As shown in Figure~\ref{KL_N_MPl},
for $M\sub{PL}=1$ TeV and $n=4$ the method gives an unbiased estimate of these
quantities, with $1\sigma$ errors of $\Delta M\sub{PL}/ M\sub{PL}\sim 15\%$,
$\Delta n\sim 0.75$ (strongly correlated).

\section{CONCLUSIONS}
If there are indeed extra dimensions large and numerous enough to reduce the
fundamental Planck scale to the TeV range, black hole production at accelerators,
with substantial cross sections, is a real possibility. Such a discovery would
arguably be the most profound advance in fundamental physics since general
relativity.  It would also constitute the first observation of Hawking
radiation.  However, as I have repeatedly emphasised, there remain many
uncertainties about the precise nature, evolution and ultimate fate of the
objects that would be formed. The strategy adopted here has been to start
with a crude but flexible simulation framework, and to look for
observables that are less sensitive to some of these uncertainties, in order
to stimulate thinking about how the fundamental parameters could be measured.
If real data on this process do become available, there will undoubtedly be an
explosion of theoretical and experimental activity that will rapidly reduce
the uncertainties.

\begin{acknowledgments}
Many thanks to members of the Cambridge SUSY Working Group, in particular
the other authors of Ref.~\cite{Harris:2004xt}, for their collaboration and
comments, and to Steve Giddings for helpful suggestions.
The hospitality of the CERN Theory Group during part of this work
is gratefully acknowledged. This work was supported in part by the U.K. Particle
Physics and Astronomy Research Council.
\end{acknowledgments}


\begin{thebibliography}{99} 

\bibitem{Giddings:2001bu}
S.~B.~Giddings and S.~D.~Thomas,
``High energy colliders as black hole factories: The end of short  distance
physics,''
Phys.\ Rev.\ D {\bf 65}, 056010 (2002)
[arXiv:hep-ph/0106219].

\bibitem{Dimopoulos:2001hw}
S.~Dimopoulos and G.~Landsberg,
``Black holes at the LHC,''
Phys.\ Rev.\ Lett.\  {\bf 87}, 161602 (2001)
[arXiv:hep-ph/0106295].

\bibitem{Antoniadis:1990ew}
I.~Antoniadis,
``A Possible New Dimension At A Few Tev,''
Phys.\ Lett.\ B {\bf 246}, 377 (1990);
N.~Arkani-Hamed, S.~Dimopoulos and G.~R.~Dvali,
``The hierarchy problem and new dimensions at a millimeter,''
Phys.\ Lett.\ B {\bf 429}, 263 (1998)
[arXiv:hep-ph/9803315];
I.~Antoniadis, N.~Arkani-Hamed, S.~Dimopoulos and G.~R.~Dvali,
``New dimensions at a millimeter to a Fermi and superstrings at a TeV,''
Phys.\ Lett.\ B {\bf 436}, 257 (1998)
[arXiv:hep-ph/9804398].

\bibitem{Randall:1999ee}
L.~Randall and R.~Sundrum,
``A large mass hierarchy from a small extra dimension,''
Phys.\ Rev.\ Lett.\  {\bf 83}, 3370 (1999)
[arXiv:hep-ph/9905221].

\bibitem{Yoshino:2002tx}
H.~Yoshino and Y.~Nambu,
``Black hole formation in the grazing collision of high-energy particles,''
Phys.\ Rev.\ D {\bf 67}, 024009 (2003)
[arXiv:gr-qc/0209003].

\bibitem{Eardley:2002re}
D.~M.~Eardley and S.~B.~Giddings,
``Classical black hole production in high-energy collisions,''
Phys.\ Rev.\ D {\bf 66}, 044011 (2002)
[arXiv:gr-qc/0201034].

\bibitem{Yoshino:2005hi}
H.~Yoshino and V.~S.~Rychkov,
``Improved analysis of black hole formation in high-energy particle
collisions,''
Phys.\ Rev.\ D {\bf 71}, 104028 (2005)
[arXiv:hep-th/0503171].

\bibitem{Harris:2004xt}
C.~M.~Harris, M.~J.~Palmer, M.~A.~Parker, P.~Richardson, A.~Sabetfakhri and B.~R.~Webber,
``Exploring higher dimensional black holes at the Large Hadron Collider,''
JHEP {\bf 0505}, 053 (2005)
[arXiv:hep-ph/0411022].

\bibitem{Harris:2004mf}
C.~M.~Harris,
``Physics beyond the standard model: Exotic leptons and black holes at  future
colliders,'' Cambridge Ph.D.\ thesis,
arXiv:hep-ph/0502005.

\bibitem{Ida:2002ez}
D.~Ida, K.~y.~Oda and S.~C.~Park,
``Rotating black holes at future colliders: Greybody factors for brane
fields,''
Phys.\ Rev.\ D {\bf 67}, 064025 (2003)
[Erratum-ibid.\ D {\bf 69}, 049901 (2004)]
[arXiv:hep-th/0212108].

\bibitem{Hannestad:2001xi}
S.~Hannestad and G.~G.~Raffelt,
``Stringent neutron-star limits on large extra dimensions,''
Phys.\ Rev.\ Lett.\  {\bf 88}, 071301 (2002)
[arXiv:hep-ph/0110067];
``Supernova and neutron-star limits on large extra dimensions reexamined,''
Phys.\ Rev.\ D {\bf 67}, 125008 (2003)
[Erratum-ibid.\ D {\bf 69}, 029901 (2004)]
[arXiv:hep-ph/0304029].

\bibitem{Dimopoulos:2001qe}
S.~Dimopoulos and R.~Emparan,
``String balls at the LHC and beyond,''
Phys.\ Lett.\ B {\bf 526}, 393 (2002)
[arXiv:hep-ph/0108060].

\bibitem{Koch:2005ks}
B.~Koch, M.~Bleicher and S.~Hossenfelder,
``Black hole remnants at the LHC,''
arXiv:hep-ph/0507138.

\bibitem{Anchordoqui:2003ug}
L.~A.~Anchordoqui, J.~L.~Feng, H.~Goldberg and A.~D.~Shapere,
``Inelastic black hole production and large extra dimensions,''
Phys.\ Lett.\ B {\bf 594}, 363 (2004)
[arXiv:hep-ph/0311365].

\bibitem{Cardoso:2005jq}
V.~Cardoso, E.~Berti and M.~Cavaglia,
``What we (don't) know about black hole formation in high-energy collisions,''
Class.\ Quant.\ Grav.\  {\bf 22}, L61 (2005)
[arXiv:hep-ph/0505125].

\bibitem{Horowitz:2005rs}
G.~T.~Horowitz,
``Higher dimensional generalizations of the Kerr black hole,''
arXiv:gr-qc/0507080.

\bibitem{Harris:2005jx}
C.~M.~Harris and P.~Kanti,
``Hawking radiation from a (4+n)-dimensional rotating black hole,''
arXiv:hep-th/0503010;

\bibitem{Duffy:2005ns}
G.~Duffy, C.~Harris, P.~Kanti and E.~Winstanley,
``Brane decay of a (4+n)-dimensional rotating black hole: Spin-0 particles,''
JHEP {\bf 0509}, 049 (2005)
[arXiv:hep-th/0507274].

\bibitem{Casals:2005sa}
M.~Casals, P.~Kanti and E.~Winstanley,
``Brane decay of a (4+n)-dimensional rotating black hole. II: Spin-1 particles,''
arXiv:hep-th/0511163.

\bibitem{Ida:2006tf}
  D.~Ida, K.~y.~Oda and S.~C.~Park,
  ``Rotating black holes at future colliders. III: Determination of black hole
  evolution,''
  arXiv:hep-th/0602188.

\bibitem{Kanti:2002nr}
  P.~Kanti and J.~March-Russell,
  ``Calculable corrections to brane black hole decay. I: The scalar case,''
  Phys.\ Rev.\ D {\bf 66}, 024023 (2002)
  [arXiv:hep-ph/0203223];
 ``II: Greybody factors for spin 1/2 and 1,''
  Phys.\ Rev.\ D {\bf 67}, 104019 (2003)
  [arXiv:hep-ph/0212199];

\bibitem{Harris:2003eg}
C.~M.~Harris and P.~Kanti,
``Hawking radiation from a (4+n)-dimensional black hole: Exact results  for the
Schwarzschild phase,''
JHEP {\bf 0310}, 014 (2003)
[arXiv:hep-ph/0309054].

\bibitem{Harris:2003db}
C.~M.~Harris, P.~Richardson and B.~R.~Webber,
``CHARYBDIS: A black hole event generator,''
JHEP {\bf 0308}, 033 (2003)
[arXiv:hep-ph/0307305].

\bibitem{Corcella:2000bw}
G.~Corcella {\it et al.},
``HERWIG 6: An event generator for hadron emission reactions with  interfering
gluons (including supersymmetric processes),''
JHEP {\bf 0101}, 010 (2001)
[arXiv:hep-ph/0011363];
``HERWIG 6.5 release note,''
arXiv:hep-ph/0210213.

\bibitem{Boos:2001cv}
E.~Boos {\it et al.},
``Generic user process interface for event generators,''
arXiv:hep-ph/0109068.

\bibitem{Moretti:2002eu}
S.~Moretti, K.~Odagiri, P.~Richardson, M.~H.~Seymour and B.~R.~Webber,
``Implementation of supersymmetric processes in the HERWIG event  generator,''
JHEP {\bf 0204}, 028 (2002)
[arXiv:hep-ph/0204123].

\end{thebibliography}
\end{document}